# Integration of Cloud Computing and Web2.0 Collaboration Technologies in E-Learning

Rasha Fouad AlCattan

Department of Computer Information System, King Abdul-Aziz University, Jeddah, Kingdom of Saudi Arabia

*Abstract*—**Cloud computing technology is an emerging new computing paradigm for delivering computing services. Although it still in its early stage, it has changed the way how many applications are developed and accessed. This computing approach relies on a number of existing technologies, such as Web2.0, virtualization, Service oriented architecture (SOA), Web services, etc. Cloud computing is growing rapidly and becoming an adoptable technology for the organizations especially education institutes, with its dynamic scalability and usage of virtualized resources as a service through the Internet. Today, eLearning is also becoming a very popular and powerful trend. However, in traditional web-based e-learning systems, building and maintenance are located onsite in institutions or enterprises, which causes lot of problems to appear, such as lacking the support of underlying infrastructures, which can dynamically allocate the needed calculation and storage resources for e- learning systems. As the need for e learning is increasing continuously and its necessary for e learning systems to keep pace with the right technology needed for development and improvement. However, today's technologies (such as Web 2.0, Cloud, etc.) enable to build more successful and effective educational environment, that provide collaboration and interaction in eLearning environments. The challenge is to use and integrate these technologies in order to construct tools that allow the best possible learning results. Cloud computing and Web2.0 are two areas that are starting to strongly effect how the development, deployment and usage of e- Learning application. This paper presents the benefits of using cloud computing with the integration of Web 2.0 collaboration technologies in eLearning environment.**

## Introduction

In the last decades, the rapid developments of internet and information technology made the innovation for various kinds for technology possible, the nature of the Web and the way the users accessing web resources for personal, educational, business, employment, entertainment, healthcare, and other social purpose, have been changed. Within the last 15 years, the Internet nature was constantly changing from static environment Web 1.0 to a highly dynamic media and more collaborative environment Web 2.0 that allows end users to run software applications collaborate, share information, and creates new services online [1] . Based on decades of researches, recently the term cloud commuting has emerged as a hot topic in the distributed computing community, virtualization, utility computing, and recently networking, and Web and software services. It is the new net revelation that many believe that it will reshape the IT industry.

Cloud computing involves a service-oriented architecture; minimize information technology overhead for the end-user, great flexibility, reduce total ownership cost, on-demand services, and so on. At the same time, the users will be unaware of where the resources and services are hosted and how they are delivered in the cloud environment [2]. As the growth of cloud computing is very fast, users can obtain the essential software and computing capability at a faster rate, which leads to tremendous improvements in the IT infrastructure and industries, and has become the recent movement in computing environment. There is no hesitation that the future goes to the cloud computing. This new environment supports the creation of new generation of web applications that can run on an extensive range of hardware devices, while data is stored inside the cloud.

Today, we can see that Cloud computing has been applied in many domains for many organizations such as E-commerce, health care and education especially in the ELearning environments.

E-Learning society is facing challenges in optimizing resource allocations, dealing with dynamic demands for accessing information and knowledge anywhere and anytime, dealing with rapid storage growth requirements, cost managing and flexibility, improving infrastructure and Lack of personalization. Furthermore, managing collaboration activities, communication and providing feedback to other students are most of the time difficult and time consuming. Yet, the need for e learning is increasing continuously and its necessary for e learning systems to keep pace with the right technology needed for development and improvement.

The purpose of this research will focus on the benefits of using cloud computing with the integration of Web 2.0 Collaboration technologies on the eLearning environment, particularly in terms of collaborative activities and increasing educational performance in an eLearning environment,

The research is organized as follows: section 1 and 2 gives introduction about Cloud computing history and definition. Section 3 provides an overview of the cloud architecture, its delivery services and deployment models. Section 4 introduces the benefits of cloud computing. Section 5 and 6 introduces cloud computing in r-learning environment and its architecture. Finally, introduce the benefits from applying the e-learning systems in the cloud.

## I. A Brief History of Cloud Computing: from Collaboration to the Cloud

Cloud computing or something within the cloud was invented in the late of 2007 [3]. In its model, Applications and documents are transferred from the traditional desktop platform to Internet platform to the cloud [4]. Users then can access and share their data and applications easily from a remote "Cloud" on-demand and according to their convenience, they will be charged only based on their consumption. Its all a about how cartelized storage enables collaboration and how multiple computers and users can work together to increase computing power.

As the Internet usage is growing all over the world, it appeared that there was no need to limit group collaboration to a single enterprise's network environment. Multiple users from multiple places, inside or outside the enterprise, desired to collaborate on common projects across the boundaries of the enterprises and share resources. To be able to do this, these common projects had to be housed in the "Cloud" of the Internet, and accessed from and Internet enabled location. [5]. Today people are using cloud services and storage to create, share, organize information from many different types, and not only from their computers but also from any device that is connected to the Internet such as mobile phones, I pad or portable music player.

## II. What is Cloud Computing?

### A. Definition of Cloud Computing

Although many formal definitions have been suggested in both academia and business for the Cloud computing, there are still no widely accepted definitions that attracted much attention. This situation due to several reasons:

- Cloud computing involves engineers and researchers from different backgrounds, who work on cloud computing from different point of view. e.g., Grid computing, software engineering and database.
- Technologies enabling Cloud Computing, such as Web 2.0 and Service oriented computing, are still in a growing and continuous development process.
- Computing Clouds still lack large-scale deployment and practice, which would lastly explain the main concept of Cloud computing [3].

Despite these issues, there appear common Key elements that are widely used in the Cloud Computing community, these common keys were provided by U.S. NIST (National Institute of Standards and Technology) [6] which includes cloud architectures, security, and deployment strategies: "*Cloud computing is a model for enabling convenient, on- demand network access to a shared pool of configurable computing resources (e.g., networks, servers, storage, applications, and services) that can be rapidly provisioned and released with minimal management effort or service provider interaction*".

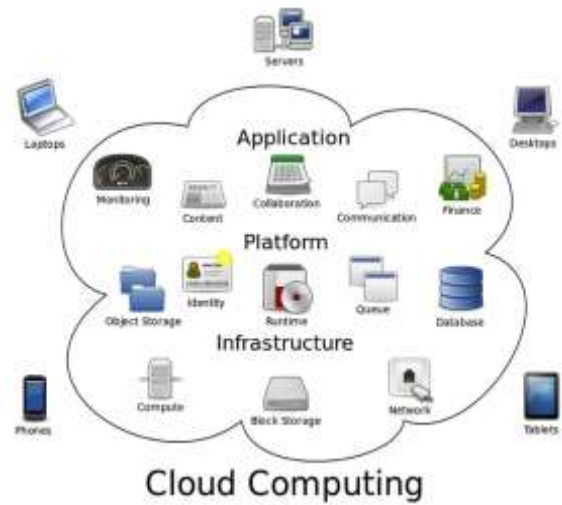

Figure 1. Cloud Computing [7]

### B. Charectaristics of Cloud Computing

There are five essential characteristics mentioned by [6] and [8] that describe the essential elements of cloud computing:

#### i. On-demand self-service, and pay-by-use

User with an instant need for resources in a particular timeslot can benefit from computing resources such as (network storage, software use, CPU time, etc.) in suitable, automatic and self-service way with out human interaction with resources providers. The self-service nature of cloud computing allows enterprises to create flexible environments that develop and bond based on the workload and target performance constraints. And the pay-by-use nature of cloud computing make the users pay only for what they used from cloud providers [4].

#### ii. Broad network access

Resources are available and can be delivered over the network (e.g. Internet) through standard mechanisms and used by mixed client platforms (e.g., laptops, mobile and PDAs) as well as any other cloud-based software services suited for user [9] [4].

#### iii. Resource pooling (shared infrastructure)

Cloud service providers and computing resources including (storage, memory, processing, virtual machines and network bandwidth) are pooled together to server multiple users across the Internet using either *multi-tenancy model* or the *virtualization model.* This means that different physical and virtual resources are dynamically assigned and reassigned according to user demand [6]. The reason for creating such a pool-based computing model is due to two important factors: *economies of scale* and *specialization*. [6]. Pool-based models results in making physical computing resources become 'invisible' or unseen to users, they do not have control or knowledge about the location, creation, and originalities of these resources (e.g. database, CPU, etc.), users have no explicit knowledge of the physical location where their data is going to be stored in the Cloud [4].

#### iv. Rapid elasticity and flexibility

Computing resources become instant rather than persistent for the users. They quickly coordinated to the real demand, quickly increasing the cloud capabilities for a service if the request rises, and quickly releasing the capabilities when the need for falls. , This automated process decreases the locating time of the new computing capabilities when they are needed. The resources appear to be unlimited and infinite to the users and the consumption can quickly rise in order to meet ultimate requirement at any time so can be purchased in any quantity at any time [10].

*v. Measured Service*

Even though computing resources are pooled and shared by multiple users, the resources usage can be automatically monitored, controlled and optimized. And this is because of the cloud infrastructure that can use suitable mechanisms to manure the usage for each individual user through metering capabilities [4] [9].

### III. HOW CLOUD COMPUTING WORKS? THE ARCHITECTURE OF CLOUD COMPUTING

There is a wide rage of solutions provided to users by cloud-based applications, to help analyzing and describing the cloud-based systems, many researchers refer to cloud solutions in term of its *service model and deployment model*. These two terms initiated by National Institute of Standards and Technology (NIST) [6].

*A. Cloud Service Delivery Model*

A cloud can interact with user or applications (client) in different ways, through what is called services. Across the web, cloud Computing has four different delivery models. They are: Infrastructure as a Service (IaaS), Platform as a Service (PaaS), and Software as a Service (SaaS) and other sub-services: [11].

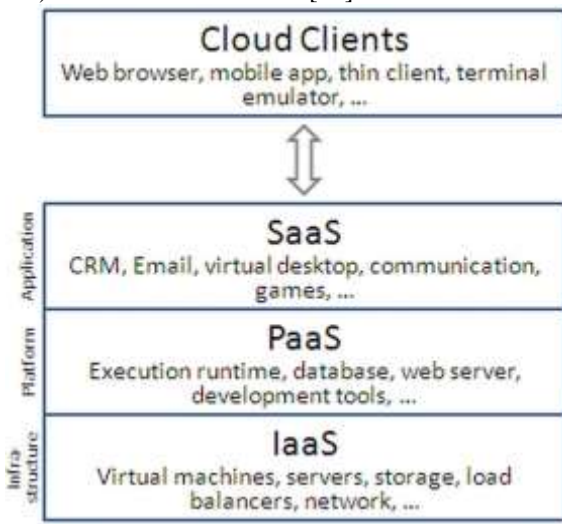

Figure 2. Cloud Service Delivery Models [7]

*i. Infrastructure as a Service (IaaS)*

This service is the foundation of all the cloud services. Cloud customers immediately use IT basic infrastructures (processing, raw storage, networks, firewalls, and other basic computing resources) provided by vendors in the IaaS cloud in a virtual platforms. Applications and resources are placed on these infrastructures thus extremely reduced massive initial investment, hardware is completely abstract and consumers use infrastructure as a service without the need to know about the underlying complexities, they can directly access resources and storage over the network. Virtualization is broadly used in IaaS cloud in order to integrate and mix physical resources to meet increasing or shortening resource demand from the customers. Virtualization essential strategy is to develop independent virtual machines (VM) that are separated from both the underlying hardware and other VMs, this strategy is not the same as multi-tenancy model, which targets to convert the application software architecture in such a way that several instances from many cloud customers can run on a single application (i.e. the same logic machine). Examples of IaaS include Private cloud, Amazon Elastic Compute Cloud (EC2), Rackspace Joyent, IBM Computing on Demand, Windows Server and System Centre and VMware [4] [9].

*ii. Platform as a Service (PaaS)*

This service model lies directly above (IaaS) on the stack, its target is not end-users, but the developers. It provides programming environments (PE) and execution environments (EE) where protective software written in a specific programming language can be executed. At this level, cloud vendors extract everything up to Operating System and middleware. Its means that vendors provide the underlying hardware technology such as: development tools and software for building application to customer, operating systems, network support and Database solutions. It offers services for each stage of software development, testing and maintenance and also sets of programming languages, which users can use to develop their own applications. Commercial examples for (PaaS) include Microsoft Windows Azure and Google App Engine [9] [12].

*iii. Software as a Service (SaaS)*

This service provides a cloud-based foundation for software and applications over the network on demand. Multiple end users or organizations can access SaaS web-delivered contents. They are available via Internet browser on a pay-as-you go basis. The advantages from SaaS service are: simplicity of integration as user only need one browser, lower cost as the data center exist within the cloud, and scalability as customer can add users to get the same benefits of commercially licensed as needed. By disregarding the demand to install and run the application on the customer's computer, SaaS eases the customer's load of software maintenance, ongoing operation, and support. Thus, it is important to know that the difference between SaaS and PaaS is that SaaS only hosts finished cloud applications whereas PaaS presents a development platform that hosts both finalized and on going cloud applications. Most widely used examples of SaaS include Gmail, Google Docs, Exchange online Business Productivity Online Suite, CRM Online, and Salesforce.com [10] [9].

*iv. Human as a Service (HuaaS)*

This service model is the upper layer of the cloud-computing stack [11]. It displays that cloud model is not limited to IT services, but can also include services provided by humans. Humans have certain abilities and skills that beat computer systems, their creativity as an important strength that appear in some tasks such as translation or design services, and their technical integration as resources is a subject of specific interest. The main subcategory in (Huaas) is *crowdsourcing* "where a group of human resources use the Internet to perform tasks of varying complexity and scope for a customer." [11]. Example for crowdsourcing is Amazon: Amazon Mechanical Turk.

   v.   *Data storage as a Service (DaaS)*

This service can be seen as a special type of (IaaS). Users via Internet can access data in different formats coming from different sources and the delivery of virtualized storage on demand becomes a separate Cloud service: "data storage service". (DaaS) allow customers to pay for their actual usage rather than the site authorization for the entire database. Additional to traditional storage interfaces such as relational database management system (RDBMS) and file systems, some (DaaS) offer table-style ideas that are intended to scale out to store and retrieve a vast quantity of data within a very limited timeframe, often too large, expensive or too slow for most commercial RDBMS to be managed. Examples of DaaS include Amazon S3, Google BigTable, and Apache HBase [4].

B.   *Cloud Application Deployment Model*

Recently, four deployment models have been defined in cloud computing community; each presents balancing benefits, and has its own trade-offs.

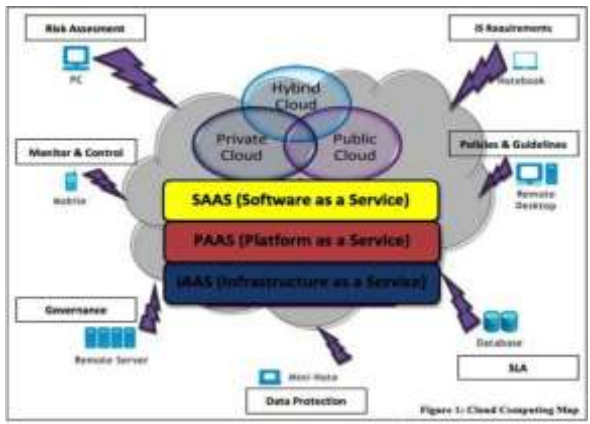

Figure 3.   Cloud Deployment Model [13]

   i.   *Private Clouds*

The cloud infrastructure is functioned exclusively within a single enterprise; applications are built, managed and controlled by the enterprise or a third party despite its location. The reasons for Private cloud within an organization are: First, maximize and optimize the usage of available resources in house. Second, security issues such as data privacy and trust. Third, cost for transferring data from local IT infrastructure to a Public Cloud is still need to be considered. Fourth, enterprises always ask for full control over serious activities that exist behind their firewalls. Finally, academics need private cloud for research and teaching reasons. [4]. Thus, Private clouds are clients built for the private use of one client, giving the highest control over data, security, and quality of service within the enterprise. They can be deployed in enterprise data center or at colocation facility [9].

   ii.   *Public cloud*

This is the most widespread of all the models and usually the less expensive solution, it is available to be used by the general public. Because of its openness, it may be owned and managed by cloud customers or a cloud service provider has the ownership of the public cloud with its own policy, value, advantage, price, and charging model. Customers and providers are most likely to be assorted together on the cloud's servers, storage systems, and networks. Public clouds are mostly of the time hosted away from the enterprise; they provide a way to reduce customer risk and cost by presenting a flexible, impermanent extension to enterprise infrastructure. The main disadvantage of the public cloud that it could be less secure because it gives extra load of confirming all accessed applications and data on the public cloud are not exposed to unwanted attacks. Many popular cloud services are public clouds including Amazon EC2, S3, Google AppEngine, and Force.com. [4].

   iii.   *Hybrid clouds*

This model provides virtual IT solutions by combining two or more clouds (private, community, or public), that keep being exclusive objects but are combined by consistent or technology that allows data and application movability (e.g., cloud bursting for load-balancing between clouds). Enterprises use the hybrid cloud model to optimize their resources and to increase their essential abilities by margining outside enterprise functions into the cloud and still controlling essential activities in the enterprise and maintain service levels in the face of rapid workload variations, through private cloud [4]. This is usually appears when using the storage clouds to support Web 2.0 applications. A hybrid cloud can also control planned workload points. Often called "surge computing," a public cloud can achieve tasks that can easily be organized on a public cloud. Some issues need to be respected, such as the relation between data and processing resources. The smaller is the data, or the displaced of application, the more successful a hybrid cloud can be than if larger amounts of data must to moved into a public cloud for a small amount of processing [8]. Hybrid Cloud provides extra secure control of the data and applications and permits different participants to reach information over the web. It has an exposed architecture that permits interfaces with other management systems.

   iv.   *Community cloud*

This model is deployed to be shared for several organizations that have a common interest. Such as government, healthcare, schools within a university. They use the same cloud infrastructure, policies, values requirements, and worries. The cloud infrastructure can be handled locally in the enterprise or by third-party and hosted internally or externally. The costs are range over fewer users than a public cloud, but more than a private cloud, thus there is a small cost savings noticed [12].

*C. The Enabeling Technology Behind Cloud Computing*

There are many enabling Technologies behind cloud computing, [3] had identified several technologies such as Virtualization technology, Worldwide distributed storage system, Web service and Service Oriented Architecture (SOA) and Web 2.0.

*i. Web 2.0*

Represents the evolution of the World Wide Web; it stands for the web applications that enable interactive information sharing, user-centered design, and collaboration on the World Wide Web. Web 2.0 is a collective term of Web-based technologies that include blogging, wikis, multimedia sharing sites, podcasting, social networks, social bookmarking sites, Really Simple Syndication (RSS) feeds, content generator and other evolving forms of participating and social media. The main concept behind Web 2.0 is to enhance the interconnectivity of Web applications and allows users to access Web in easy and efficient way. Cloud computing services essentially are Web applications that us the Internet as a computing platform and provide on demand computing services. Therefore, as a normal technical evolution, the Cloud computing adopts the Web 2.0 technique, It is believed that cloud computing is putting a fundamental infrastructure of Web 2.0; it enables and is enhanced by the Web 2.0 Model.

*ii. Virtualization technology*

The base of the cloud computing as it provides a flexible hardware services. Virtual machine techniques such as VMware provide on demand virtual IT infrastructure, also VPN a Virtual network advances, enables users to access cloud services through a customized network environment.

*iii. Web service and Service Oriented Architecture (SOA)*

Because the clouds are Web services, the services enterprise within Clouds can be handled in a Service Oriented Architecture (SOA), also the cloud services can be used in a SOA application environment, which make them available and accessed through many spread platforms across the Internet.

*iv. Worldwide distributed storage system*

First, a network storage system, (e.g. data center) used for backup and data storage by distributed storage providers. Google File System is a good example; Mashup also is a Web application that mixes data from different sources into a single combined storage tool.

Second, a distributed data system that delivers data sources accessed in a semantic way. Virtual Data System (VDS) is a good example where users can find data in a large distributed environment by logical name not the physical locations.

IV. WHY CLOUD COMPUTING REALLY MATTERS

According to [5] Cloud Computing technology is important for both developers and users for many reasons:

*A. Cloud Computing for Developers*

- Offers more amounts of storage and processing power to run their applications.
- Provides different and new ways to access information, connect people and resources from different locations world wide, process and analyze data.
- Developers feel free from the physical constraints.

*B. Cloud Computing for End Users*

- User is not restricted to single computer, location, or network.
- User can access his/her applications and documents hosted in the cloud from anywhere at any time.
- The fear of loosing data if the computer crashed is gone.
- Benefit of group collaboration, users worldwide can access, share, update, the same documents or applications in the real time. It's an entire new world of collaborative computing, all enabled by the concept of cloud computing.

V. CLOUD COMPUTING AND E-LEARNING

Today, Learners are looking forward to having a learning schedule and network-learning environment that is fixable and meet their own needs. Because of the low enrolment in onsite classes many educational institutions offer courses and some times the entire degree program through distance education or online. This new frontier of education is known as ELearning. Though it will not replace traditional education methods, but will significantly enhance the efficiency of education, update technology, tools and methods concept for education, thus the parts of teachers cannot be replaced.

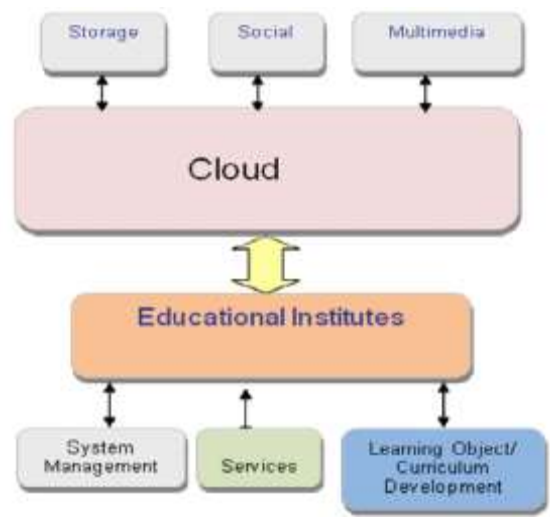

Figure 4. Modified E-learning based on Cloud [14]

*A. E-Learning*

An Internet-based learning process, using Internet technology to design, implement, choose, handle, support and extend learning with the advantages of flexibility, diversity, measurement, opening and more [14]. However, in traditional eLearning environments, services are normally based on anytime technologies that do not cope with "everyone" and "everywhere" aspects, systems building and maintenance are placed in the educational institutions, which caused a lot of problems and a lot of investments without capital gain in return [15] [14]. Moreover as mentioned by [16] eLearning society met challenges in different area including: optimizing resource allocations, dealing with dynamic needs for accessing and retrieving information from anywhere and anytime, dealing with quick storage growth requirements, cost management and flexibility, refining infrastructure, Lack of personalization where platforms available to learners have same content, and when they happen to have different learning demands or learning roles, they need to enter different learning systems which lessen learning desire and motivation. Nowadays, the need for e learning is increasing continuously and its necessary for e learning systems to keep pace with the right technology needed for development and improvement. However, today's technologies (such as Web 2.0, Cloud, etc.) enable to build more successful and effective educational environment, which deliver collaboration and interaction in eLearning environments. The challenge here is how to use and integrate these technologies to develop tools that allow the best achievable learning results [17]. Cloud computing and Web2.0 are two important technologies that are starting to strongly impact the development, deployment and usage of e-learning applications.

*B. Cloud Computing*

Provides a new way of deploying applications, as it becomes an important technology because how it deals with the resources effectively and in dynamic scalability. It provides a new way of deploying applications. In cloud, teachers will remain as leading roles and contribute in development of eLearning cloud. The blended learning strategy, interactive content and virtual collaboration should improve the education environment, we can have Infrastructure as a Service (IaaS), Platform as a Service (PaaS) or Software as a Service (SaaS) and After computing resources are virtualized, they can be afforded as services for educational institutions to be accessed by students and teachers. Cloud computing can benefit ELearning systems by using: 1- Infrastructure: an eLearning solution is on the provider's infrastructure. 2- Platform: the usage and development of an eLearning solution based on the provider's development interface. 3- Services: the eLearning solution given by the provider [18].

*C. Web 2.0*

A major technology that supports dynamic and content publishing over the Internet; it unites tagging culture and use Internet to make links and connections with information, it allows people to create, exchange, publish, and share information in a new way of communication and collaboration. Applying Web 2.0 Applications such as (Wikipedia, blogs, YouTube, social networks, RSS, tagging) to eLearning can improve interactive communication and collaboration amongst students, who have similar learning resources, or help find the resources and share them with others in the Web-based learning. As a result students become the consumers and producers of learning resources. As a result, Web 2.0 provides a learning environment that has the ability to change the basic nature of learning and teaching, by the development of learner controlled learning web [19].

*D. Integartion of Cloud Computing and Web 2.0 in E-Learning*

In the mid of last decade, ELearning has been a very active research and applied field. Various technologies have been added in order to improve and enhance eLearning systems and eLearning process. Over the recent years, the focus of changing traditional eLearning systems to more collaborative interactive learning environments has increased. (Méndez & González, 2011) Explains limitations of traditional e learning pointing that system creation and maintenance are placed inside the educational institutions, which can cause a lot of problems including: huge investments with no gain and lack of development potential.

In 2009, cloud computing is presented by [20] as a new computing model to implement eLearning ecosystem to over come the problems in the traditional system. The author believed that cloud computing is able to add some critical features to eLearning ecosystem. Such as: configuration at real-time, utilization of resources, on-demand resource sharing and better management for software or hardware. The cloud- based system, supports the construction of new generation of eLearning systems that is accessible from a wide range of hardware devices, whereas storing data inside the cloud. Ouf, Nasr and Helmy [17] had proposed an e learning system based on the integration of cloud computing and Web 2.0 technologies to meet the requirements for e learning environment such as flexibility and compliance towards students' needs and concerns and improve and enhance the efficiency of learning environment. The authors had mentioned the most important cloud-based services such as Google App Engine and classified the advantages from implementing cloud-based ELearning 2.0 applications such as scalability, feasibility and availability, also emphasized the improvements in cost and risk management.

Similarly Zheng [16] proposed a platform architecture based on the integration of cloud computing and web 2.0 for developing intelligent virtual learning community and make the learning environment more productive, scalable, flexible and adjustable towards students' demands and needed information and communication technologies. The architectures was based on Windows Azure, which is a cloud-computing platform and infrastructure and Web 2.0 Mashups was adopted to integrate Web Feed and Web API to create ELearning system on local computers which made it easy for students to access the system with different clients including different browsers, RSS reader, and personal mobile phone. The author Usage of cloud computing and web2.0 for e- learning affects the way an eLearning software projects are managed, the proposed intelligent virtual learning community enhanced the efficiency of learning environment, provide up-to-date resources, constancy, quality of service guaranteed, dependability, scalability, minimize time, efficient usage of resources, flexibility, and maintaining of eLearning system.

Chandral [21] presented a Hybrid Instructional Model as a mix of traditional classroom and online learning and how it has been customized for running e-learning applications on the cloud infrastructure. The author focused on the current r-learning architecture models and the issues in its applications, in particular its openness, scalability and development costs. E-learning systems are not dynamically designed and sometimes they are difficult to expand, also integration with other e-learning systems is costly. The author in this study suggested the hybrid cloud delivery model that can help in fixing this issue.

There are many cloud computing service providers that offer support for learning systems and enable students to benefit from the provided tools and applications in the cloud such as email, file storage, collaborative tools, create and sharing contents Google is one of the famous vendors for a cloud computing service, in 2008 Google-App engine was released as platform for users to build and host applications, and Google-App for education as software service with Google applications, GoogleDoc, collaborative tools, email and file storage all in the cloud. Microsoft also, announced Windows Azure in 2010, Azure is a flexible platform that let users solve their educational, by addressing their needs and roved educational tools. IBM also offered CloudBurst, a prepackaged cloud includes hardware, software, application and middle ware for faster application development [17].

## VI. E-LEARNING CLOUD BASED ARCHITECTURE

Many ELearning cloud architecture had been introduced and demonstrated by many researchers and according to [22] they are generally similar in terms of the different layers that they consist of and their functions. Such cloud based eLearning architecture is presented by [14], [20]and [23]. The architecture can be divided into the following layers: Infrastructure layer, software resource layer, resource management layer, service layer, and application layer. (Figure 5)

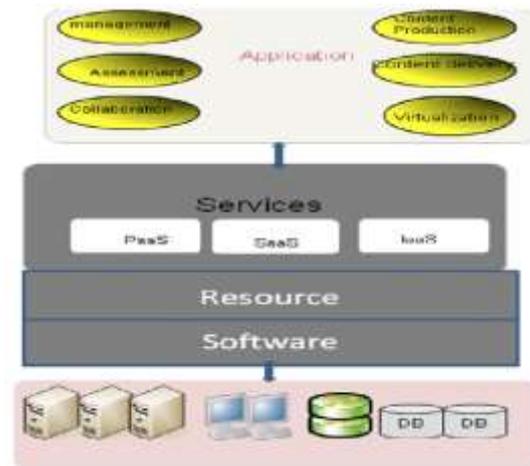

Figure 5.   E-learning Cloud Architecture [14]

### A. Infrastructure layer

Works as a dynamic and scalable physical host point, it is placed in the bottom layer of cloud service and composed of information infrastructure including (Internet/Intranet, system software, information management system and some known software and hardware) and teaching resources that are gathered in traditional teaching model and distributed in various departments and domain. This layer provides the basic computing power such as physical and CPU memory. The use of virtualization technology enables upper software platform to call physical server, storage and network form virtualization group.

### B. Software resource layer

Offers a unified interface for eLearning developers, by using middleware technology; different software resources are combined to deliver a unified interface for developers to make it easy for them to develop applications based on available software resources and make them available for users in the cloud.

### C. Resource management layer

Attains free combination of software and hardware resources. The integration of virtualization and cloud computing enables achieving scheduled strategy, on-demand flow and distribution of software over different hardware resources.

### D. Service layer

Have three levels of cloud computing services, SaaS (Software as a service), PaaS (Platform as a service), and IaaS (Infrastructure as a service). In SaaS, customers can access with a service monthly fee via the Internet, with no need to purchase software and hardware, and no need to maintain and upgrade.

### E. Application layer

The applications of the teaching resources integration in the cloud-computing model, includes interactive courses and sharing teaching resources. Interactive courses, which can be more effective than traditional teaching, are mainly for the teachers, they take advantage of the underlying information resources, and the course content and the progress can adjust anytime according to the feedback. Sharing of teaching resources include material, information resources and human resources sharing. This layer provides content production and delivery technology, collaborative learning, assessment and management features.

VII. BENIETS FROM APPLYING CLOUD COMPUTING TO E-LEARNING

A lot of benefits can be achieved when implementing the eLearning systems in the cloud, these benefits include [18]:

A. *Lower Cost*

The cost of building educational information system can be reduced when using cloud [24]. In addition, the software licensing cost is reduced as it is offered as a service from providers who will also take the responsibility of some computing maintenance from IT staff [22]. Students in E-learning environments do not require specific computers with large memory to store data and run the applications they need, they can run the applications from the cloud through their personal devices such as mobile, ipad and tablets. Thus, Organizations will pay per use, which is cheaper for them.

B. *Improved learning performance*

Since all the applications on the cloud, when the client machines work they will not cause any issues on the overall and learning performance. The learning process will be clearly affected in a positive way as more IT team will focus on providing better support for learners rather IT maintenance issues [25].

C. *Immediate software updates*

Students will have instant update as the applications on the cloud are automatically updated in the cloud source.

D. *Enhanced document format compatibility*

Students will not face the problem of not opining their files from different devices due to different formats compatibility, since they open files from the cloud. Thus students do not have to worry about their PCs or mobiles supported file formats.

E. *Benefits for students*

Students will have more advantages through cloud based eLearning by taking online courses, having exams online, having feedback about the courses from teachers, uploading projects and assignments online, and collaborate and share resources and course contents over the cloud. Moreover, students will have access 24/7 to up-to-date resources with all the required tools to achieve the learning goal in a flexible environment; the can resources and course contents can be evolved collaboratively over the cloud and shared.

F. *Benefits for teachers*

Teachers can prepare online tests for students, use content management to create better content resources for students, evaluate tests and projects done by students, communicate with students and send feedback [18].

G. *Data security*

Although it's obvious that there is a huge concern about the data security, as they are placed on a remote server and can be crashed with no warnings. However, cloud computing provides main security advantages for persons and organizations who use or develop E-learning environments.

H. *Better learning recourses management*

Cloud computing provides better learning resource management and better integration and consumption of learning resources and this is by providing improved management method, automated deployment and high level virtualization [22]. Additionally, It also supports the use of multimedia learning contents in mobile learning and offers a chance to build a mobile educational resource library [26].

VIII. CONCLUSION AND FUTURE WORK

Cloud computing, as a new development Internet-based computing model, is a significant alternative for today's educational perspective, especially in the eLearning environment. Students and teachers have the chance to rapidly and economically access several application platforms and resources across the web pages on-demand anytime anywhere, which result in minimizing the cost of organizational payments and presents strong functional capabilities. Recently, the need for e learning is increasing continuously yet eLearning society faced challenges in optimizing resource allocations, dealing with dynamic demands for accessing information anywhere and anytime, dealing with rapid storage growth requirements, cost managing and flexibility, refining infrastructure, Lack of personalization where the available platforms to learners are the same, thus when they have different learning demands and roles, they have to enter different learning systems which minimize their learning desire.

As a result, the need for e learning is increasing continuously and its necessary for e learning systems to keep pace with the right technology needed for development and improvement. Therefore eLearning systems must keep pace with the right technology needed for development and improvement, thus it cannot ignore the cloud computing and Web 2.0 trends, and the benefits from their integration. Using cloud computing with the integration with Web 2.0 collaboration technology for eLearning affects the way an eLearning software projects are managed, enable to build more successful and effective educational environment, that provide collaboration and interaction in eLearning environments.

The most significant benefits of cloud eLearning are that it enables resources accessibility from multiple devices, such as computers and smartphones, allows for a wide network of individuals for formal and informal learning, increase collaboration activities and enhance educational performance for learners.

In summary, the movement towards cloud computing can be a greater variation of Internet ready devices, applications accessed directly from the Web, data placed in the cloud, and company applications controlled and hosted by third party service suppliers. Our paper suggests that introducing cloud computing with the integration of Wen 2.0 collaboration technologies into eLearning environments is feasible and it can greatly increase the collaboration activities and educational performance for learners.

For future work, our aim is to implement an eLearning system by using integration of Cloud computing and Web 3.0.

ACKNOWLEDGMENT

First and foremost, I am very grateful to Allah for helping me in completing this research Special thanks and appreciation goes to my supervisor Professor. Muazzam Siddiqui for his guidance and support through the whole process.